\documentclass[aps,prb,twocolumn,showpacs,superscriptaddress,citeautoscript,longbibliography]{revtex4-1}
\usepackage{graphicx} 
\usepackage{amsmath} 
\usepackage{bm} 
\usepackage{natbib}
\usepackage{natmove}
\usepackage[colorlinks=true, linkcolor=blue, citecolor=blue, urlcolor=blue, linktoc=page, bookmarks=false, pdfstartview={FitH}, pdfborder={0 0 0.0 [3 3]}]{hyperref} 
\usepackage{color} 
\usepackage{dcolumn} 
\newcolumntype{.}{D{.}{.}{2.1}}
\newcolumntype{-}{D{.}{.}{4.0}}
\usepackage{makecell}
\usepackage{multirow}
\usepackage{cleveref}
\crefname{figure}{Fig.}{Figs}
\crefname{table}{Table}{Tables}
\renewcommand{\today}{\number\day \space \ifcase \month \or January\or February\or March\or April\or May \or June\or July\or August\or September\or October\or November\or December\fi \space \number\year} 
\def\m1r{\multicolumn{1}{r}}

\begin{document}

\title{Itinerant Ferromagnetism in $p$-doped Monolayers of MoS$_\text{2}$}
\author{Yuqiang Gao}
\email[Email: ]{Y.Gao@utwente.nl}
\affiliation{Faculty of Science and Technology and MESA$^+$ Institute for Nanotechnology, University of Twente, P.O.~Box~217, 7500~AE Enschede, The Netherlands}
\affiliation{Department of Applied Physics, Northwestern Polytechnical University, Xi'an, China}

\author{Nirmal Ganguli}
\email[Email: ]{nganguli@iiserb.ac.in}
\affiliation{Faculty of Science and Technology and MESA$^+$ Institute for Nanotechnology, University of Twente, P.O.~Box~217, 7500~AE Enschede, The Netherlands}
\altaffiliation{Present address: Department of Physics, Indian Institute of Science Education and Research Bhopal, Bhauri, Bhopal 462066, India}

\author{Paul J. Kelly\thanks{corresponding author}}
\email[Email: ]{P.J.Kelly@utwente.nl}
\affiliation{Faculty of Science and Technology and MESA$^+$ Institute for Nanotechnology, University of Twente, P.O.~Box~217, 7500~AE Enschede, The Netherlands}
\affiliation{The Center for Advanced Quantum Studies and Department of Physics, Beijing Normal University, 100875 Beijing, China}
\date{\today}
\begin{abstract}
Density functional theory is used to explore the possibility of inducing impurity band ferromagnetism in monolayers of semiconducting MoS$_2$ by introducing holes into the narrow Mo $4d$ band that forms the top of the valence band. A large out of plane anisotropy is found for unpaired spins bound to the substitutional acceptor impurities V, Nb and Ta that couple ferromagnetically for all but the shortest separations. Using the separation dependent exchange interactions as input to Monte Carlo calculations, we estimate ordering temperatures as a function of the impurity concentration. For about 9\% of V impurities, Curie temperatures in excess of 160 K are predicted. The singlet formation at short separations that limits the ordering temperature is explained and we suggest how it can be circumvented.
\end{abstract}

\maketitle

\begin{figure}[b]
\includegraphics[scale = 0.35]{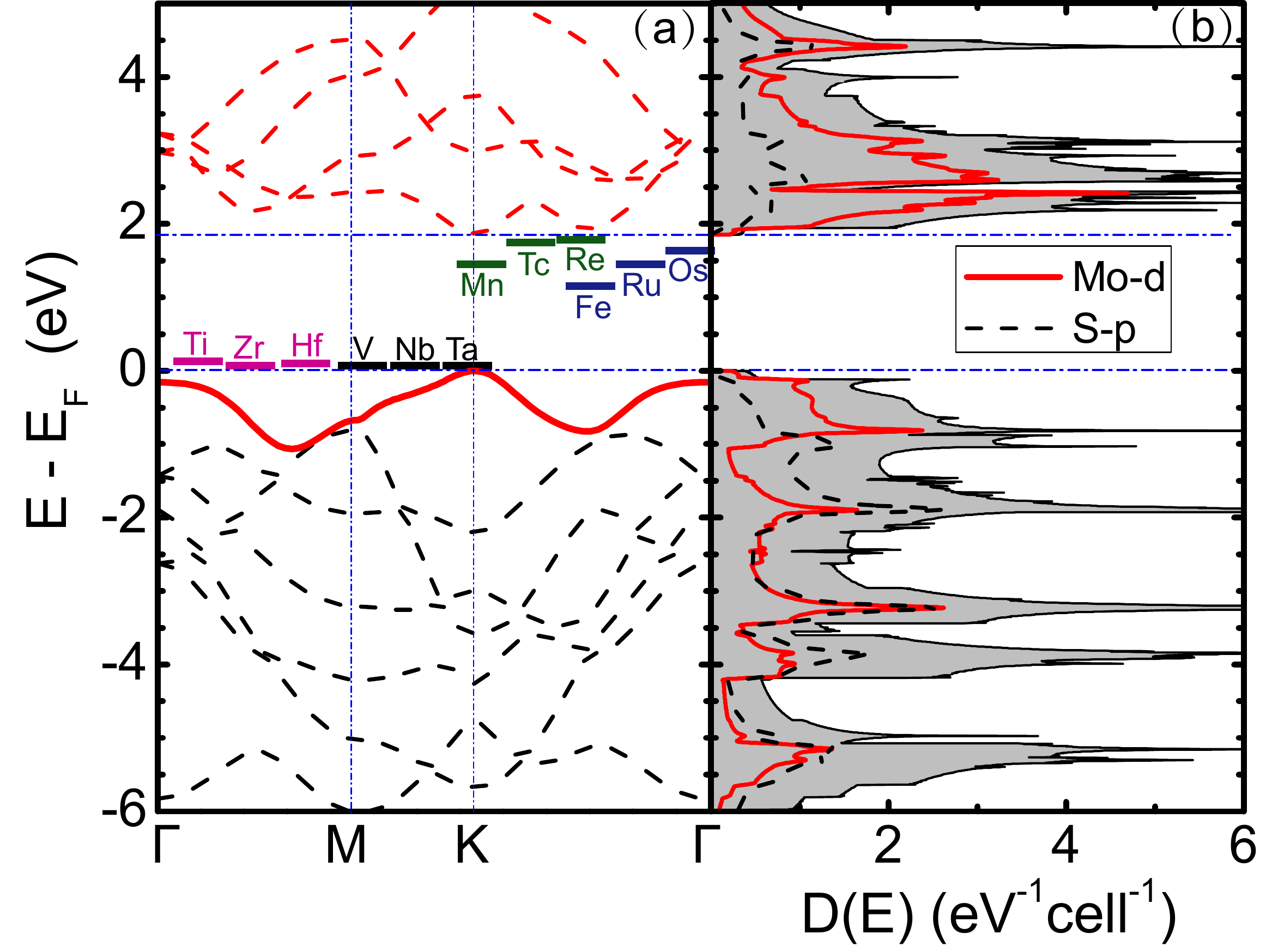}
\caption{\label{fig:bandDos}(a) The band structure of an MoS$_2$ monolayer and (b) the corresponding total, and Mo-$d$ and S-$p$ projected densities of states.  The spin-degenerate (point) defect levels that result from substituting Mo with various transition metals are indicated: group IV (pink), group V (black), group VII (green), and group VIII (blue).}
\end{figure}

{\color{red}\it Introduction.---}
The extraordinary interest sparked by the discovery of  intrinsic ferromagnetism in the two dimensional van der Waals semiconducting crystals CrGeTe$_3$ \cite{Gong:nat17} and CrI$_3$ \cite{Huang:nat17} has led us to examine the possibility of inducing ferromagnetism in monolayers of MoS$_2$ by doping the narrow Mo $d$ band that forms the top of the valence band with holes.  Theoretical analysis of the MX$_2$ layered transition metal dichalcogenides (M = Mo, W; X = S, Se, Te) more than forty years ago \cite{Bromley:jpc72, Mattheiss:prb73} revealed the curious electronic structure shown in Fig.~\ref{fig:bandDos}. The $d$ valence states of the Mo atoms interact with the chalcogen $p$ states to form a substantial band gap leaving a single ``nonbonding'' Mo $d$ band (lhs, solid red line) in the hybridization gap between bonding states (dashed black lines, with nominal X $p$ character) and antibonding states (dashed red lines, with nominal Mo $d$ character). The projected densities of states (DoS) in Fig.~\ref{fig:bandDos}(b) show the considerable mixing that actually occurs. The reduced coordination number of metal atoms in two-dimensional structures leads to smaller bandwidths and higher state densities than in three dimensions making this system favourable for the occurrence of itinerant ferromagnetism. Motivated by predictions of high-temperature ferromagnetism in narrow impurity bands \cite{Edwards:jpcm06}, we examine the behaviour of single acceptor states in the low concentration regime. We show that Mo$_{1-x}$V$_x$S$_2$ monolayers should become ferromagnetic semiconductors with Curie temperatures much larger than those found for either CrGeTe$_3$ or CrI$_3$.

Monolayers of MoS$_2$ were among the first two-dimensional (2D) materials to be prepared by micromechanical exfoliation \cite{Novoselov:pnas05}. Renewed interest in this otherwise well known semiconductor was rekindled by the subsequent observation that the monolayer bandgaps were direct \cite{Splendiani:nanol10} and substantially larger than for the bulk \cite{Mak:prl10}, and by the realization of transistors with large on/off ratios \cite{Radisavljevic:natn11}. A direct band gap makes MX$_2$ semiconductors promising candidates for optoelectronic applications. The lack of inversion symmetry in monolayer transition metal dichalcogenides (TMD) and large spin orbit splitting of the degenerate $K$ and $K'$ valence band maxima is promising for the subject of ``valleytronics'' whereby carriers in  different valleys can be manipulated using magnetic fields, magnetic substrates and optical pumping \cite{Mak:natn12, Smolenski:prx16, Zhao:natn17}. The exchange interaction introduced by a magnetic dopant can break the time reversal symmetry and lead to valley polarization \cite{Cheng:prb14}. According to the Mermin-Wagner theorem \cite{Mermin:prl66}, there is no long range ferromagnetic ordering in a strictly 2D isotropic Heisenberg model. However, the theorem is invalidated by magnetic anisotropy and many observations of magnetic ordering in thin layers of metallic ferromagnets as well as the recent discoveries \cite{Gong:nat17, Huang:nat17} show that this frequently happens in practice.

Doping MX$_2$ monolayers has been considered theoretically with various defects and substitutions leading to the formation of local moments \cite{Ataca:jpcc11, Cheng:prb13, Yue:pla13, Ramasubramaniam:prb13, Mishra:prb13, Dolui:prb13, Lu:nrl14, Yun:pccp14, Andriotis:prb14, Qi:jpcm14, Gil:jpcm14, Miao:jms16, Robertson:acsn16, Zhao:jac16, Singh:am17, Miao:ass18}. However, no attempt has been made to calculate  magnetic anisotropies without which there is no long-range ferromagnetism and only a few attempts have been made to study the separation dependence of the exchange interaction \cite{Ramasubramaniam:prb13, Mishra:prb13, Miao:ass18}. 
The coupling between impurities is determined by the range of the localized impurity states; the deeper the levels are, the shorter the range of the effective interaction and the lower the ordering temperature for a given concentration of dopant. The short range of the effective exchange interaction leads to percolation and requires such heavy doping to achieve coherent magnetic ordering \cite{Sato:rmp10} that the final material is no longer a semiconductor. 
Motivated by the electronic structure shown in Fig.~\ref{fig:bandDos} and the promise of high temperature ferromagnetism in narrow impurity bands \cite{Edwards:jpcm06}, we adopt a different approach to making a magnetic semiconductor out of MoS$_2$. By introducing a low concentration of holes into the top of the narrow valence band, we explore the possibility of realizing a material that is ferromagnetic while remaining a semiconductor.

{\color{red}\it Method.---}
Total energy calculations and structural optimizations were carried out within the framework of density functional theory (DFT) using the projector augmented wave (PAW) method \cite{Blochl:prb94b} and a plane-wave basis  with a cut-off energy of 400 eV as implemented in the {\scshape vasp} code \cite{Kresse:prb93a, Kresse:prb96, Kresse:prb99}. We use the local spin density approximation (LSDA) as parameterized by Perdew and Zunger \cite{Perdew:prb81} to describe  exchange and correlation effects because it accurately reproduces the experimentally observed ordering of the valence band maxima \cite{Jin:prl13}. Monolayers of MoS$_2$ periodically repeated in the $c$ direction were separated by more than 20~\AA\ of vacuum to minimize the interaction. Substitutional impurities and impurity pairs were modeled using $N \times N$ in-plane supercells with $N$ as large as 15; interactions between pairs of impurities were studied in $12 \times 12$ supercells to reduce interactions between periodic images to an acceptable level. Atomic positions were relaxed using a $2 \times 2 \times 1$ $\Gamma$-centered $k$-point mesh until the forces on each ion were smaller than 0.01 eV/\AA. Spin-polarized calculations were performed with a denser mesh corresponding to $4 \times 4$ $k$-points for a $12 \times 12$ unit cell.

{\color{red}\it Single impurity limit.---}
We begin by substituting a single Mo atom in a $12 \times 12$ MoS$_2$ supercell with 3$d$, 4$d$ or 5$d$ atoms that have 1 or 2 valence electrons more or less than Mo. The energy of this 432 atom supercell is first minimized with respect to the atomic coordinates. The resulting energy levels found in the gap without spin polarization for the 12 dopant atoms considered is summarized in Fig.~\ref{fig:bandDos}. All 3$d$, 4$d$ and 5$d$ ions with one or two electrons fewer than Mo introduce shallow acceptor states as measured by the proximity of the lowest unoccupied or partially occupied electron level to the top of the valence band. Re and Tc form shallow donors but the other four elements Mn, Fe, Ru and Os give rise to deep levels and will not be considered further here. The magnetic moments obtained when spin polarization is included are given in \cref{tab:magMomMoS2}. The single shallow acceptors V, Nb and Ta and donors Tc and Re are found to polarize completely while the double acceptors and double donors (but not Fe) are spin compensated. 

The formation energy \cite{footnote1} of MoS$_2$:V is a modest 0.4 eV, while those of Nb and Ta are 0.01 eV and -0.12 eV, respectively, indicating that it should not be very difficult to substitutionally dope MoS$_2$ with these $4+$ ions \cite{Robertson:acsn16, Suh:nanol14, Chua:acscat16} that have a $d^1$ configuration, a single unpaired spin.
Under the local $D_{3h}$ symmetry of a Mo atom, the Coulomb potential of a substitutional single acceptor leads to the formation of a doubly degenerate $e'$ state with $d_{x^2-y^2}$ and $d_{xy}$ character bound to the $K$ and $K'$ valence band maxima (VBM) with a Bohr radius of $\sim 8\,$\AA\ as well as an $a'_1$ state bound to the slightly lower lying $\Gamma$ point VBM with $d_{3z^2-r^2}$ character and a Bohr radius of $\sim 4\,$\AA. In supercell calculations for single V substitutional impurities, these three levels form partially filled overlapping states. The results for the magnetic moments shown in \cref{tab:sizeVaryMagMom} can be understood in terms of how these bands shift with respect to one another and broaden as the separation between impurities decreases for smaller supercells. Discrepancies with earlier theoretical studies can be understood in terms of (i) these supercell-size dependent results and (ii) use of the GGA approximation that results in the the $\Gamma$-point VBM being too high \cite{Jin:prl13} and the hole occupying the $a'_1$ bound state \cite{Cheng:prb13, Yue:pla13, Lu:nrl14, Yun:pccp14, Andriotis:prb14, Miao:jms16, Robertson:acsn16, Zhao:jac16, Miao:ass18}.



\begin{table} [t]
\caption{\label{tab:magMomMoS2}
Calculated magnetic moments in $\mu_B$ for monolayers of MoS$_2$ doped with various transition metal atoms substituting a single Mo atom in a $12 \times 12$ supercell.}
\begin{ruledtabular}
\begin{tabular}{lccccc}
       & \multicolumn{2}{c}{Acceptors} &  & \multicolumn{2}{c}{Donors}\\
\cline{2-3}\cline{5-6}
 Group & 4               & 5               & 6         & 7              & 8  \\
\hline
3$d$   & \text{Ti}: 0.00 &  \text{V}: 1.00 & \text{Cr} &\text{Mn}: 1.00 & \text{Fe}: 2.00 \\
4$d$   & \text{Zr}: 0.00 & \text{Nb}: 1.00 & \bf{Mo}   &\text{Tc}: 1.00 & \text{Ru}: 0.00 \\
5$d$   & \text{Hf}: 0.00 & \text{Ta}: 1.00 & \text{W}  &\text{Re}: 1.00 & \text{Os}: 0.00 \\
\end{tabular}
\end{ruledtabular}
\label{tableI}
\end{table}

\begin{table}[b]
\caption{
\label{tab:sizeVaryMagMom}Magnetic moment (in $\mu_B$) calculated for an MoS$_2$ monolayer supercell doped with V, Nb and Ta as a function of the $N \times N$ supercell size without (U) and with (R) relaxation.
}
\begin{ruledtabular}
\begin{tabular}{ll...........}
$N$ &  & \m1r{4} & \m1r{5} & \m1r{6} & \m1r{7} & \m1r{8} & \m1r{9} & \m1r{10} & \m1r{11} & \m1r{12} \\
\hline
V  & \text{U} & 0.00 & 0.61 & 0.82 & 0.97 & 1.00 & 1.00 & 1.00 & 1.00 & 1.00 \\
   & \text{R} & 0.00 & 0.81 & 1.00 & 1.00 & 1.00 & 1.00 & 1.00 & 1.00 & 1.00 \\
\hline
Nb & \text{U} & 0.00 & 0.00 & 0.34 & 0.54 & 0.68 & 0.87 & 0.93 & 1.00 & 1.00 \\
   & \text{R} & 0.00 & 0.00 & 0.66 & 0.75 & 0.89 & 1.00 & 1.00 & 1.00 & 1.00 \\
\hline
Ta & \text{U} & 0.00 & 0.00 & 0.32 & 0.42 & 0.57 & 0.86 & 0.95 & 1.00 & 1.00 \\
   & \text{R} & 0.00 & 0.00 & 0.68 & 0.83 & 0.92 & 1.00 & 1.00 & 1.00 & 1.00 \\
\end{tabular}
\end{ruledtabular}
\end{table}

\begin{figure}[b]
\includegraphics[scale = 0.30]{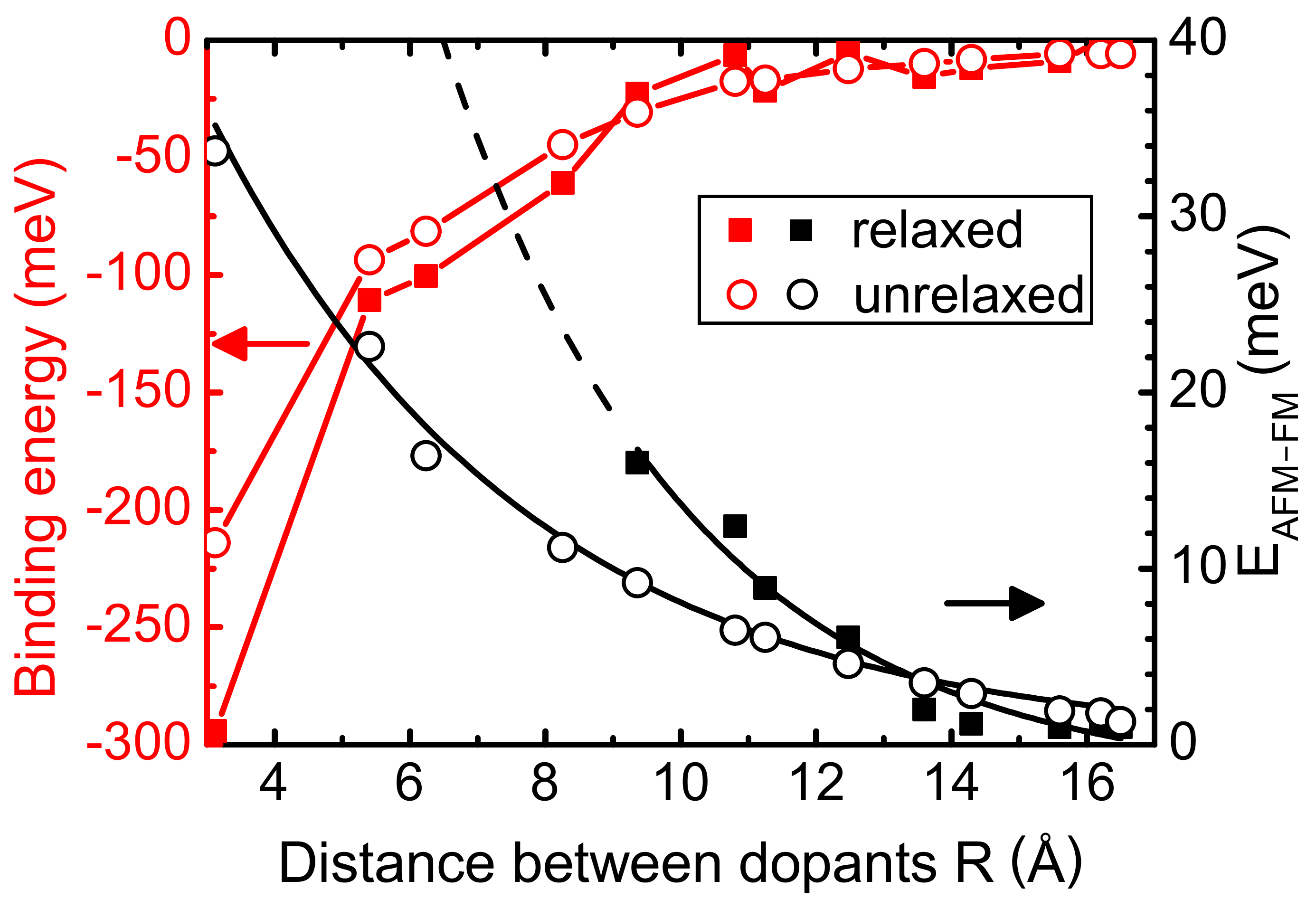}
\caption{\label{fig:bindingExchange}The difference between the total energies of antiferromagnetic and ferromagnetic states (black) and the pair binding energy (red) for pairs of V substitutional impurities plotted as a function of their separation for unrelaxed (open symbols) and fully relaxed (filled symbols) configurations in $12 \times 12$ MoS$_2$ supercells. The dashed black curve extrapolates the relaxed results to separations where quenching occurs.
}
\end{figure}

{\color{red}\it Exchange interaction.---}
We estimate the exchange interaction between pairs of V, Nb and Ta dopant atoms by calculating the total energies for substitutional pairs as a function of their separation with their magnetic moments aligned parallel (``ferromagnetically'', FM) and antiparallel (``antiferromagnetically'', AFM) in  $12 \times 12$ supercells, with and without atomic relaxation. With one dopant atom at the origin, we explore all inequivalent sites in the $12 \times 12$ supercell with the second dopant atom. The binding energy of V pairs is the energy difference $E_b(R) = E(R) - E(R=\infty)$ where $E(R)$ is the LDA total energy for a supercell with two dopants separated by a distance $R$. $E_b$ is shown as a function of $R$ in \cref{fig:bindingExchange} (left-hand side axis), along with the energy difference between antiferromagnetic and ferromagnetic states (right-hand side axis). It has a maximum magnitude of $\sim 0.3$~eV for V dopants on nearest neighbor lattice sites and decays essentially monotonically as a function of the separation. Because it is so small, it can be assumed that dopant atoms will be randomly distributed in real materials that are not fully equilibrated.

Two unpaired spins usually form a singlet state to maximize their bonding energy. We find that unrelaxed pairs of dopant atoms couple ferromagnetically for all separations. V pairs  are more strongly coupled than Nb pairs that are more strongly coupled than Ta pairs. When relaxation is included, the magnetic moment is quenched for atoms closer than a critical separation, in the case of V this is $\sim 9.4$~\AA, see  \cref{fig:bindingExchange}.  As expected for hydrogenic defect states that are more localized by a stronger central cell potential, the exchange interaction decays faster with increasing separation for V than for Nb than for Ta. To a good approximation the interaction strength only depends on the separation and decays exponentially with increasing separation with decay lengths of 3.6, 5.2 and 5.8 \AA, respectively.

According to the Mermin-Wagner theorem \cite{Mermin:prl66}, isotropic Heisenberg exchange will not yield a finite ordering temperature in two dimensions. However, MX$_2$ monolayers do not have inversion symmetry and spin orbit coupling (SOC) leads to a substantial splitting of the Kramers degenerate states at $K$ and $K'$ with $d_{xy}/d_{x^2-y^2}$ character \cite{Zhu:prb11}. In the single impurity limit, we find a 132 meV SOC-induced splitting of the $e'$ level that results in a large single ion magnetic anisotropy (SIA) of about 5~meV with preferred orientation of the magnetic moment perpendicular to the monolayer plane. The SIA is much larger than the value reported for 2D CrI$_3$ that exhibits Ising behavior \cite{Xu:npjcm18}. 

\begin{figure}[b]
\includegraphics[scale = 0.33]{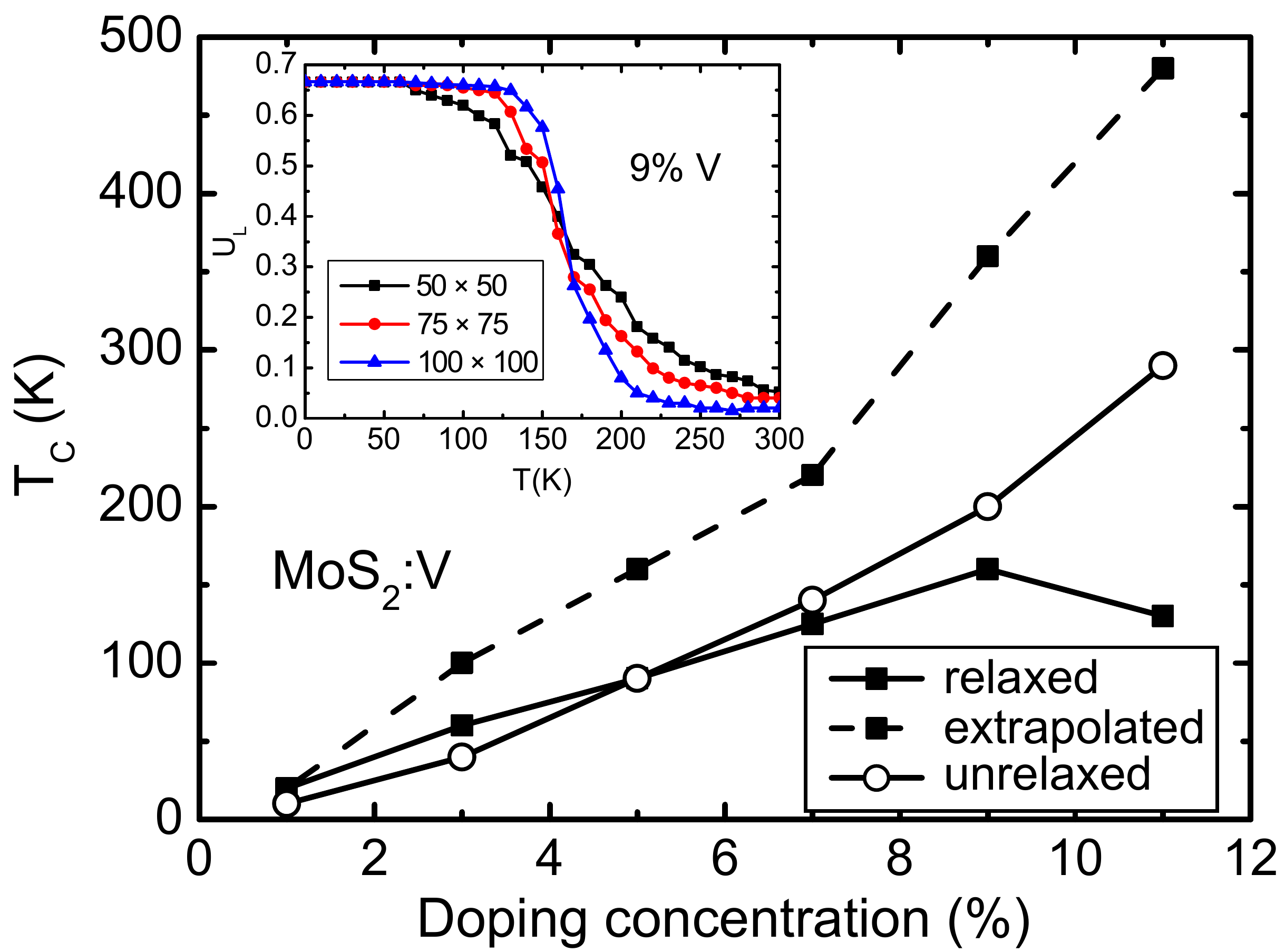}
\caption{\label{fig:dopConcTc}Variation of the ferromagnetic Curie temperature as a function of the doping concentration calculated using Binder's cumulant method and the exchange interactions shown in \cref{fig:bindingExchange} for an MoS$_2$ monolayer doped with V. The dashed curve was calculated by using extrapolated relaxed exchange interactions at separations where quenching occurs. Inset: fourth order cumulant calculated as a function of the temperature for three different lattice sizes.
}
\end{figure}

{\color{red}\it Curie temperature.---}
In the case of very strong SIA, the system can be described by an Ising model yielding a magnetically ordered phase at finite temperature \cite{Onsager:pr44, Yang:pr52}. We treat the V (Nb and Ta) doped MoS$_2$ monolayer as an Ising spin system and combine the exchange interactions calculated above with Monte Carlo calculations to estimate ferromagnetic Curie temperatures $T_C$ with Binder's cumulant method \cite{Binder:zfpb81, Landau:09}. The fourth order cumulant of the magnetization $m$, $U_L(T) = 1 - \langle m^4 \rangle / 3 \langle m^2 \rangle^2 $, is calculated for three different lattice sizes $L=50, 75, 100$ as a function of the temperature $T$ and the ordering temperature is given by the size-independent universal fixed point where the $U_L(T)$ curves intersect for different lattices sizes $L$; see inset \cref{fig:dopConcTc}.  

The resulting values of $T_C(x)$ are shown in \cref{fig:dopConcTc} for V doping concentrations $x$ in the range from 1\% to 11\%. $T_C$ initially increases with increasing doping, reaches a maximum value of $\sim$160~K (Nb and Ta are lower) for a concentration of $\sim 9\%$ before decreasing again for larger values of $x$ (squares, solid line). For the exchange interaction calculated without relaxation, there is no maximum and $T_C$ increases monotonically with concentration (circles, solid line). If we extrapolate the relaxed exchange interaction to close separations where quenching occurs (dashed line in \cref{fig:bindingExchange}), we find the ordering temperatures shown as a dashed line in \cref{fig:dopConcTc}. The behavior at close separations is limiting the maximum Curie temperature attainable making it important to understand the nature of the quenching. We examine this for the case of V.

\begin{figure}[t]
\includegraphics[scale = 0.34]{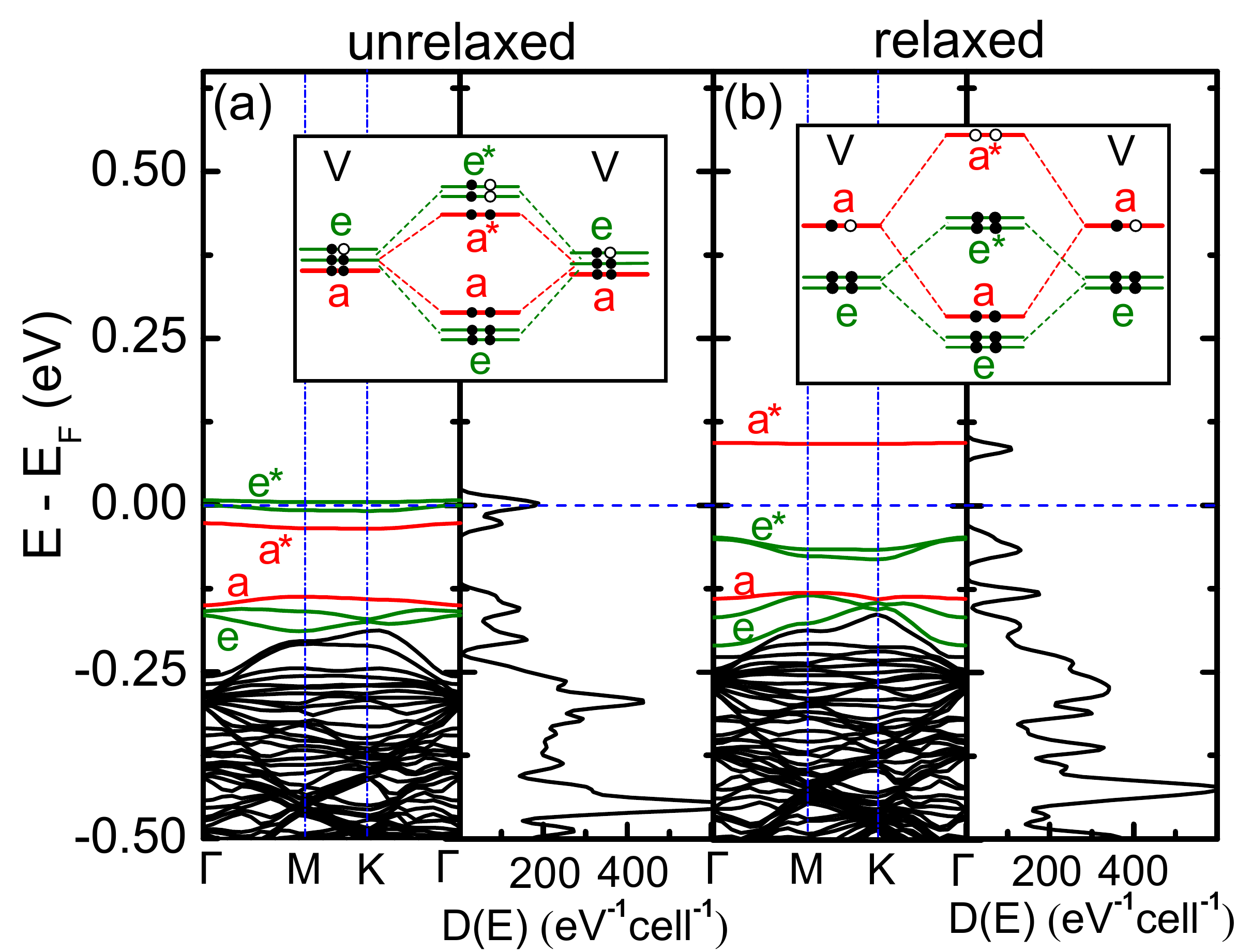}
\caption{\label{fig:band}
Unpolarized electronic structure for a $12 \times 12$ MoS$_2$ supercell with two V atoms on nearest neighbour Mo sites. Band structure and DoS for the system without (a) and with (b) atomic relaxation. Defect bands are highlighted in red and green.The insets show the coupling scheme of two adjacent VS$_6$ units with unrelaxed (left) and relaxed atomic positions (right)}
\end{figure}

{\color{red}\it Quenching of moments for close dopant pairs.---}
The magnetic moments of V dopant atoms couple ferromagnetically at separations larger than 9.4~\AA\ and are quenched at closer separations, \cref{fig:bindingExchange}. In the absence of relaxation, however, we find that V atoms couple ferromagnetically for all separations. To understand the role of the local atomic relaxation in quenching the FM coupling, we  consider a $12 \times 12$ supercell of MoS$_2$ doped with a pair of V atoms on neighboring Mo sites. The spin unpolarized supercell band structures and DoS with (right) and without (left) relaxation are shown in \cref{fig:band}.
$a'_1$ states with $d_{3z^2-r^2}$ orbital character on the neighbouring V atoms form a $\pi$ bond (red bands) while the orbitally degenerate $e'$ states with $d_{xy}/d_{x^2-y^2}$ character form $\sigma$ bonds (green bands). Without relaxation, the $\pi$ bond is weaker than the $\sigma$ bond (lhs) and this leads to holes occupying the doubly degenerate $e^*$ antibonding states, \cref{fig:band}(a) inset. In the periodic supercell, this leads to a peak in the DoS at the Fermi level and a magnetic instability.

Relaxation results in a structure where the S atoms move closer to the V atoms and the two V atoms move slightly apart. The reduced V-S bond length strengthens the $\pi$ bond by the increased hybridization between V $d_{3z^2-r^2}$ and S $p$ orbitals. The $\sigma$ bonds formed by the V {$d_{xy}$,$d_{x^2-y^2}$} orbitals are weakened by the increased V-V separation; compare the $a-a^*$ ($\Delta_\pi$) and $e-e^*$ splittings in \cref*{fig:band}(a) and \cref*{fig:band}(b). Now all states are fully filled or empty, both holes occupy the $a^*$ antibonding singlet spin state and there is no magnetic instability. 
This situation is energetically favourable as long as the energy gain from bonding ($\Delta_\pi$ for one hole on each dopant) is larger than the energy gain from exchange splitting. The exponential decrease of $\Delta_\pi$ with increasing dopant separation and almost constant exchange splitting \cite{Gunnarsson:jpf76, Janak:prb77, Andersen:physbc77a} results in quenching of the magnetic moment at close separations while a triplet state forms when the bonding interaction becomes weaker than the exchange splitting at larger separations.

{\color{red}\it Discussion.---}
The quenching of ferromagnetic pairing for close impurity pairs can be avoided by considering the host semiconductors MoSe$_2$ or MoTe$_2$ (WSe$_2$ or WTe$_2$) for which the $\Gamma$ point VBM drops with respect to the $K$-$K'$ VBM as S$\rightarrow$Se$\rightarrow$Te. Preliminary calculations show that the $a'_1$ defect levels follow the $\Gamma$ point VBM leaving the holes in the orbitally degenerate $e'$ derived impurity bands. Double acceptors would be expected to have larger magnetic moments and exchange splittings (but also be more localized and more susceptible to Jahn-Teller distortions). The MX$_2$ system offers many possibilities to tune the magnetic properties of electron and hole doped monolayers by varying the composition of the host system with M=Cr, Mo, W and X=S, Se, Te or alloys of these constitutents on either the M or X sublattice. 

As the impurity concentration is increased, the impurity levels will overlap to form narrow bands that broaden and eventually overlap the narrow Mo band that forms the top of the valence band. For itinerant electrons occupying narrow bands, it has been argued that the effective interaction predicted by the Stoner criterion will not be reduced by correlation effects or spin wave excitations \cite{Edwards:jpcm06}. For the 9\% V dopant concentration for which $T_C$ is maximum, we find an $e'$ impurity band width of $\sim 400 \,$meV. The $a'_1$ band is even narrower, only about a third as wide. Both exceed the 90 meV exchange splitting we find for single V impurities that would imply partial quenching of the magnetic moments. For the ordered V dopants studied in \cref{tab:sizeVaryMagMom}, this quenching occurs as the concentration is increased above 3\% and is complete by 6\%. In this context, we note that our LDA results provide a lower bound on the exchange interaction and ordering temperature which would be enhanced with a small Coulomb U in LDA+U calculations; a value of $U=1\,$eV is sufficient to make a $3\times3$ system ferromagnetic.

{\color{red}\it Conclusions.---}
Although the maximum value of $T_C$ we find is below room temperature, the MX$_2$ material system offers many possibilities to tune both host and dopant (electrons as well as holes) properties. The observation of strong room-temperature ferromagnetism in VSe$_2$ \cite{Bonilla:natn18} confirms that the MX$_2$ system may host interesting new magnetic effects. Very recently, there have been intriguing reports of room-temperature ferromagnetism in monolayers of WSe$_2$ \cite{Yun:arXiv18} and MoTe$_2$ \cite{Coelho:aem19} lightly doped with vanadium. 

This work was financially supported by the ``Nederlandse Organisatie voor Wetenschappelijk Onderzoek'' (NWO) through the research programme of the former ``Stichting voor Fundamenteel Onderzoek der Materie,'' (NWO-I, formerly FOM) and through the use of supercomputer facilities of NWO ``Exacte Wetenschappen'' (Physical Sciences). Y. G. thanks the China Scholarship Council for financial support. N.G.\ thanks Dr.\ Supravat Dey for fruitful discussions.

%

\end{document}